\documentclass[%
 reprint,
 aps,prd,groupedaddress
 amsmath,amssymb,
 aps,
]{revtex4-2}

\usepackage{graphicx}
\usepackage{dcolumn}
\usepackage{bm}
\usepackage{amsmath}
\usepackage{xcolor}
\usepackage{hyperref}

\newcommand{\aap}{A\&A}

\newcommand{\mnras}{MNRAS}

\newcommand{\physrep}{Phys. Rep.}

\newcommand{\jcap}{JCAP}
\newcommand{\apss}{ApSS}

\newcommand\apjl{{ ApJL}}%

\newcommand{\dHybridR}{{\it dHybridR}}
\newcommand{\orcid}[1]{\hspace{1mm}\href{https://orcid.org/#1}{\includegraphics[height=0.3cm,keepaspectratio]{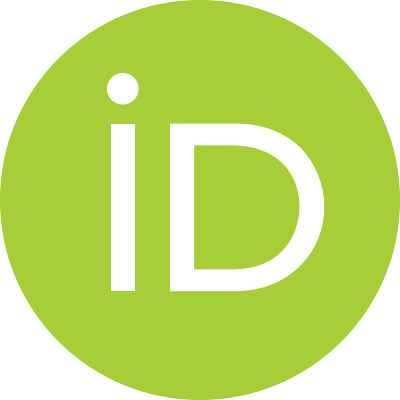}}}
\begin{document}

\preprint{APS/123-QED}

\title{Investigating Nonlinear Landau Damping in Hybrid Simulations}

\author{Benedikt Schroer\orcid{0000-0002-4273-9896}}
\email{bschroer@uchicago.edu}
\affiliation{Department of Astronomy and Astrophysics, University of Chicago, 5640 S Ellis Ave, Chicago, IL 60637, USA}

\author{Damiano Caprioli\orcid{0000-0003-0939-8775}} 
\affiliation{Department of Astronomy \& Astrophysics, University of Chicago, Chicago, IL 60637, USA}
\affiliation{Enrico Fermi Institute, The University of Chicago, Chicago, IL 60637, USA}

\author{Pasquale Blasi\orcid{0000-0003-2480-599X}}
\affiliation{Gran Sasso Science Institute (GSSI), Viale Francesco Crispi 7, 67100 L'Aquila, Italy}
\affiliation{INFN-Laboratori Nazionali del Gran Sasso (LNGS), 
via Giovanni Acitelli 22, 67100 Assergi (AQ), Italy}

\date{\today}

\begin{abstract}
Phenomenological studies of cosmic-ray self-confinement often hinge on the linear theory for the growth rate of the streaming instability and for the damping rate of the generated magnetic modes. 
Largely different expressions exist, especially for the rate of nonlinear Landau damping, which is often assumed to be the most important damping mechanism in warm ionized plasmas.
Using hybrid-PIC simulations in the resonant streaming instability regime, we present a comprehensive assessment of nonlinear Landau damping and show that the damping rate at a given scale depends on the power in magnetic fields on larger scales.
Furthermore, we find that an inverse cascade develops, which produces magnetic fields on scales larger than the resonant ones. 
Here we extend previous results obtained for a mono-energetic distribution of non-thermal particles to the case of broader CR distributions, as a first step towards developing phenomenological models.
Pre-existing turbulence of Alfv\'enic nature at large scales severely affects the damping of waves produced by low-energy CRs; 
depending on its amplitude, such a turbulence may inhibit the growth of streaming instability so that CRs are either self-confined at all energies or not at all.
\end{abstract}

\maketitle

\section{Introduction}
\label{sec:intro}

The transport of cosmic rays (CRs) is a highly complex problem, as they diffuse through different phases of the interstellar medium (ISM) and feed back on their own transport properties in a highly nonlinear way via the excitation of streaming instabilities.
Recent measurements of secondary-to-primary CR flux ratios \cite{ams21,dampe22,CALET22} reveal several spectral breaks, some of which might be related to the nature of this nonlinear transport on Galactic scales in the hot ionized medium \cite{blasi+12,chernyshov+24,chernyshov+22} typical of the extended Galactic halo, where CRs are expected to spend most of their time.
In this picture, CRs might be responsible for producing their own scattering centers via the excitation of the resonant streaming instability \cite{kulsrud+68, skilling75b}.
Typically, the growth of this instability is balanced by scale-dependent damping mechanisms \cite{kulsrud+71} that shape the energy dependence of the transport process, although many other processes are at work \cite{hopkins+21,hopkins+22b, lemmerz+24, Cerri2024}. 
In the hot ionized medium, the dominant damping mechanism is thought to be nonlinear Landau damping (NLLD) \cite{lee+73,volk+82,nava+19,blasi+12,chernyshov+24,armillotta+24,evoli+18b}.

Due to the complexity of the problem, it is impossible to study it entirely from first principles. Hence, phenomenological studies typically rely on descriptions of the corresponding damping rate based on the linear analysis of interaction of two waves interacting in a high-$\beta$ plasma \cite{lee+73}.
When two parallel-propagating waves of slightly different wavelengths interact, there are two different situations to consider: If the waves have \textit{different polarizations}, both waves lose energy and heat the background plasma. 
If they share the \textit{same polarization}, the wave with the smaller wavelength loses energy, while the other gains energy.
The results of the linear analysis are then extended to a spectrum of waves \cite{volk+82}, which has led to two very different prescriptions in the literature for the damping rate to be used:
\begin{enumerate}
    \item In one prescription, the damping rate $\Gamma_{\rm NLLD}$ at a given wavenumber $k$ is proportional to the magnetic field on larger scales 
    \begin{equation}\label{eq:integral}
        \Gamma_{\rm NLLD}\propto k \int_0^k\mathrm{d}k' \delta B(k')^2 \,.
    \end{equation}
    This formulation captures the result that modes get damped because of the interaction with all of the larger-wavelength waves. 
    It was applied, for instance, to studies of Galactic CR transport \cite{chernyshov+24} and Galaxy formation \cite{thomas+19}.
    \item Alternatively, it was postulated that only waves close in wavenumber space interact \cite{volk+82}. In this formulation the damping rate is approximately given by the magnetic fields around the considered wavenumber 
    \begin{equation}\label{eq:resonant}
        \Gamma_{\rm NLLD}\propto k^2\,\delta B(k)^2 \,.
    \end{equation} The resulting damping rate was adopted in the investigation of galaxy clusters \cite{wiener+18}, heating in the ISM \cite{wiener+13}, launching Galactic winds \cite{armillotta+24}, Galactic CR transport \cite{blasi+12} and CR escape from their sources \cite{recchia+16,recchia+22,nava+19,dangelo+16}.
\end{enumerate}

The difference between the two cases depends mainly on the spectrum of the perturbations, which for self-generated transport is determined by the spectrum of particles. 
This problem was discussed by \cite{lagage+83b} in the case of particle acceleration, where the authors argued that the second recipe could be used safely since the spectrum of accelerated particles is relatively hard. In general, it is true that these descriptions may lead to largely different results in terms of energy dependence of the diffusion coefficient, for instance during transport in the Galaxy. 
This is particularly important for the interpretation of both the Galactic diffuse gamma-ray emission \cite{lhaaso23} and the TeV halos around sources \cite{hawc17_coll,lhaaso+21}, since these studies rely on extrapolating the diffusion coefficient inferred from B/C to higher energies. 
Understanding this phenomenon is also important for the CR scattering at energies $\gtrsim 1$ TeV, where self-generation may be ineffective.

In order to get physical insights into this problem, we perform numerical simulations of the CR resonant streaming instability to self-consistently assess the production and damping of magnetic fluctuations, and use the results to inform phenomenological investigations of CR transport.
Introducing the plasma $\beta$ as the ratio of the ion thermal velocity and the Alfvén velocity, $\beta =2 (v_{th}/v_A)^2$, we note that previous numerical attempts mainly focused on the low-$\beta$ regime, in which NLLD is negligible or on cases where ion-neutral damping is important, employing particle-in-cell (PIC) \cite{holcomb+19} or magnetohydrodynamics (MHD)+PIC \cite{bai+19,plotnikov+21} simulations.
In $\beta\ll 1$ simulations, damping is suppressed and hence, the instability saturates once the CRs are scattered efficiently to drift collectively with the Alfvén speed \cite{holcomb+19}.

In a recent study \cite{schroer+25}, we used hybrid PIC simulations (with kinetic ions and fluid electrons) to investigate CR transport in self-generated fields under the influence of NLLD in a $\beta\gtrsim 1$ plasma. 
The rate of NLLD was found to depend on the power in magnetic fields on larger scales, confirming the integral scaling in Eq. \ref{eq:integral}. 
An inverse cascade developed, moving power of magnetic fields from small (resonant) to large scales, compatible with findings of PIC simulations \cite{shalaby+21}.
Such a cascade can be understood as interactions between waves with the same polarization, where the larger-wavelength waves gain energy, as described above.
The study was performed with a monoenergetic CR population.
In order to extend these findings to cases of phenomenological interest, it is important to test what happens when CRs with different energies are present in the system and test the numerical limitations of these simulations.
In this article, we present the results of simulations with multiple CR species of different energies, generating the resonant streaming instability in a $\beta\gg1$ plasma. 
We show that the effect of the presence of the new CR species can be understood in the context of a consistent theory. 
Furthermore, we demonstrate that in a box that is periodic for the waves, the inverse cascade moves power to larger and larger scales and a balance between damping and growth at large scales is impossible. 
As shown in section~\ref{sec:results}, this finding is in agreement with the growth rate of the inverse cascade and the damping rate of NLLD based on the linear two-wave analysis \cite{lee+73}.
By demonstrating that the behavior of the spectrum of waves is compatible with the naive expectations based on the linear analysis, we pave the way to apply our results to phenomenological models, in particular by discussing in section \ref{sec:implications} the potential role of large-scale, pre-existing, Alfvènic turbulence on the self-confinement of low-energy CRs.

The article is structured as follows: In section~\ref{sec:setup}, we describe our simulation setup. Our results are presented in section~\ref{sec:results} and in section~\ref{sec:conclusions} we conclude.

\section{Hybrid PIC simulations}
\label{sec:setup}
We study the self-consistent, self-generated diffusion of CRs using \dHybridR{}, a relativistic hybrid code with kinetic ions and (massless, charge-neutralizing) fluid electrons \citep{gargate+07,haggerty+19a}. 
We consider here an adiabatic closure for the electron pressure, i.e., $P_e\propto \rho^{5/3}$ \citep{caprioli+18}.
\dHybridR{} is suited to properly simulate both the non-resonant and resonant streaming instability \citep{haggerty+19a,schroer+21,schroer+22,zacharegkas+24}
and should capture all the relevant mechanisms for saturation of the resonant streaming instability, i.e., nonlinear Landau damping \cite[also see][]{kunz+14a}.
By choosing a hybrid code we are able to simulate longer time scales compared to PIC codes, which is essential for exploring low growth rates of the resonant streaming instability.

In our simulations, velocities, time and length scales are normalized to the Alfvén speed $v_A$, the ion cyclotron time $\Omega_0$ and the ion skin depth $d_i= v_A/\Omega_{0}$.
Magnetic fields are normalized to the background magnetic field $B_0$.
Particle number densities are normalized to the initial one of the background plasma $n_0$.

Our simulation box is quasi-1D along $x$ and retains all three components of the momenta and electromagnetic fields.
The total size is $60000\,d_i \times 5\,d_i$, divided into $120000 \times 10$ cells. 
The speed of light is fixed to $c=100\,v_A$ and the background magnetic field is directed along the $x$-axis with strength $B_0$.
The background plasma is periodic in all directions and the background plasma is sampled from a Maxwellian with a plasma beta of $\beta=2 v_{th,i}^2/v_A^2=2$ and $128$ particles per cell, in order to reduce the magnetic field noise floor.
CRs have periodic conditions in $y$, open boundaries in $x$ and we use $16$ particles per cell.
We keep injecting CRs with a distribution that is isotropic in their rest frame and then boosted into the simulation frame with momentum $p_{bst}$, which leads to a mildly super-Alfvénic drift speed in the background plasma (for more details see Ref.~\cite{schroer+25}).
Depending on the run, we inject one or both of the following CR distributions:
A population with $p_{iso}=100\,m v_A$ (Lorentz factor $\sqrt{2}$) and $n_{CR} = 1.8\times 10^{-4}\,n_0$ and $p_{bst}=5\, m v_A$, which leads to a drift speed of $4\,v_A$ or
a CR species with $p_{iso}=300\,m v_A$ and $n_{CR}=2\times 10^{-5}\,n_0$ and $p_{bst}=18\, m v_A$, which leads to a drift speed of $12\,v_A$.
The parameters of the second population are chosen to mimic CRs in the Galaxy, i.e., a drift speed that increases with momentum and a power-law spectrum.
The time step is $0.005\,\Omega_{ci}^{-1}$.

In total we run 4 different cases: $\mathcal{A}$. The first CR population and open boundary conditions for the fields; $\mathcal{B}$. The first CR population and periodic boundary conditions for the fields; $\mathcal{C}$. The second CR population and open boundary conditions for the fields; $\mathcal{D}$. Both CR populations and open boundary conditions for the fields.
For case $\mathcal{C}$ and $\mathcal{D}$, we increase the simulation size to $180000\,d_i \times 5\,d_i$, divided into $360000 \times 10$ cells, to resolve the larger scattering lengths of the higher energy particles. 

\section{Results}
\label{sec:results}
\subsection{The Benchmark Case $\mathcal{A}$}
The whole evolution of the CRs and magnetic field of case $\mathcal{A}$ is described extensively in Ref.~\cite{schroer+25}.
Hence, here, we will only briefly summarize these results to lay the foundation for the comparison with the other cases and refer the reader to Ref.~\cite{schroer+25} for more details.

The evolution can be divided into four stages.
In Stage I, CRs stream through the box without scattering and excite the resonant streaming instability with a rate consistent with the analytical growth rate \cite{amato+09}.
After a few e-folds, the magnetic field is large enough to scatter the CRs and they slow down and transition to a diffusing population.
At the same time, the growth rate of the instability slows down, while magnetic perturbations start to grow on scales larger than the CR gyroradius $R$, i.e., nonresonant scales, due to an inverse cascade.
Entering Stage III, the growth of the instability halts due to damping and
the CR drift speed slowly increases as the magnetic field shifts towards nonresonant, large scales.
As the total energy in magnetic fields grows, an increasing amount of energy gets advected out of the box due to it being open for the Alfvén waves.
Approximately one Alfvén-crossing time after the magnetic field on resonant scales peaked, the box is filled with waves generated by the inverse cascade and a steady state is reached.
This final stage indicates that a balance is reached between the growth of new fields and the waves leaving the box. 

\subsection{Case $\mathcal{B}$ and the Effect of Boundary Conditions}
Since the damping rate at a given $k$ depends on the fields at larger scales, the growth of large-scale modes can only be balanced by damping after creating enough power at even larger scales. Hence, in the previous case, a steady state is only reached due to waves being able to leave the box.
Otherwise, waves would accumulate at larger and larger scales to balance the growth of smaller scale waves.
To further test this picture, we repeat the simulation of case $\mathcal{A}$, but with periodic boundaries for the waves.

The evolution of both, the CRs and the magnetic field of case $\mathcal{B}$ is illustrated in Fig.~\ref{fig:caseII}.
Similarly as before, we can identify four different phases in the evolution that indicate different interplay of CRs and fields.
The first and second stage are identical to their equivalents in case $\mathcal{A}$ and seem rather universal: There is always a short, initial period, in which CRs do not scatter and freely stream, followed by a period of increased scattering and resulting slowdown of CRs.
Scattering is accompanied by a reduction in growth rate from the unperturbed growth rate of our initial, pristine CR distribution to a new growth rate given by the self-scattered distribution.

After $\sim 20$ e-folds, Stage III starts, in which CRs reach a diffusive regime for an extended amount of time.
Although the magnetic field behaves similar to case $\mathcal{A}$ (flattening of $B_\perp$ on resonant scales and growth on large scales) the CR drift speed remains constant instead of increasing with time.

During Stage III, the magnetic field contained on all scales that are able to resonantly scatter CRs, plateaus, despite $v_D>v_A$.
Stage III is the final result that one would have expected based on scenarios in which residual growth is compensated by damping processes, as described in sec.~\ref{sec:intro}.
It corresponds to the physical picture that is invoked for our Galaxy, that CRs generate their own scattering centers and then enter a diffusive regime with possible drift speeds larger than $v_A$ because the residual growth rate is canceled by damping processes.
However, around $t\,\gamma_{max}\approx 50$ we observe a fourth regime in our periodic box.

Contrary to case $\mathcal{A}$, there is no efficient way for the system to get rid of the energy in magnetic fields at large scales.
Hence, the total field continues to increase with time, with an increasing fraction of it accumulating at scales larger than $R$.
Stage IV starts when the power contained on scales larger than the resonant scales becomes comparable with the power at resonant scales.
The reduction in drift speed hints at a change of scattering regime from resonant scattering to scattering off large-scale turbulent magnetic fields that dominate the power spectrum.
As expected, a steady state cannot be achieved.
Without a process to balance the wave growth, the field keeps increasing and cascading to larger scales.

\subsubsection{Magnetic Field Cascade}
Such a behavior is consistent with the expectation based on the linear analysis of two-wave interactions and its extrapolation to spectra of waves \citep{lee+73,volk+82}: the damping rate at a given scale $k$ is given by the integral over all the power in magnetic fields at larger scales, see Eq.~\ref{eq:integral}. Following the same argument, the growth due to the inverse cascade is given by the same integral but over all scales smaller than the given scale and of the same polarization $\Gamma_g \propto  k\int_k^\infty\mathrm{d}k' \delta B(k')^2$. As a result, the growth due to the inverse cascade can only be canceled by damping at a single wavenumber $k_0$ where there is equal power above and below that scale. 
At all $k\leq k_0$, the growth due to interactions with smaller wavelength modes is stronger than damping resulting in net growth ($\Gamma_g\geq \Gamma_{\rm NLLD}$). On the other hand, at all $k\geq k_0$, the damping will take over and magnetic fields are damped ($\Gamma_{\rm NLLD}\geq \Gamma_g$). 

\begin{figure}
    \centering
    \includegraphics[width=0.48\textwidth, trim=12 12 10 10, clip=true]{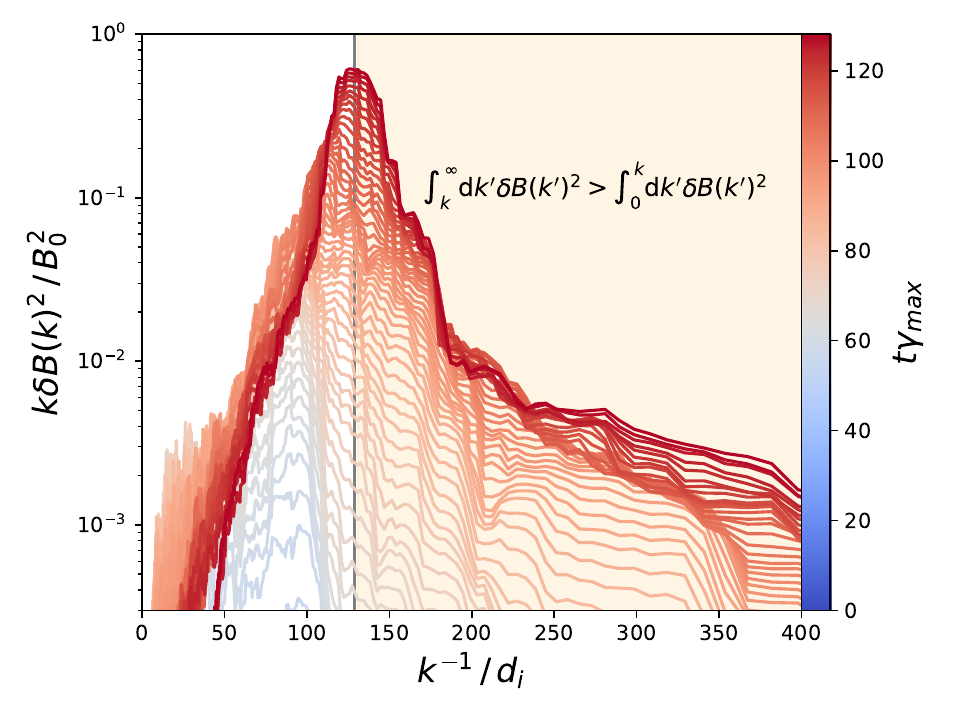}
    \caption{Time evolution (color coded) of the power in left-handed modes for case $\mathcal{B}$. The vertical line indicates the scale $k_0^{-1}$ for which the power contained on smaller scales is equal to the power in larger scales. In the yellow-shaded region the growth due to the inverse cascade dominates over damping.}
    \label{fig:caseII_power}
\end{figure}

At a fixed $k$, $\Gamma_{\rm NLLD}$ is increasing with time because of the growth of large-scale modes and $\Gamma_g$ is decreasing since some small-scale modes are damped away. Hence, $k_0$ evolves with time and at late times, the power spectrum can be seen to follow this trend, as illustrated in Fig.~\ref{fig:caseII_power}. First, growth is centered around the maximally growing wavelength of the resonant streaming instability $k_{res}^{-1}\sim 100\,d_i$.
Around $t \gamma_{max}\sim 70$, the growth due to the inverse cascade at larger scales is larger than the growth at $k_{res}^{-1}$ due to the streaming instability and the cascade. As a result, the power spectrum becomes very peaked around $k_0$ with a peak that slowly moves towards larger scales, consistent with the above behavior. At the end of the simulation, the peak shifted from $k_{res}^{-1}$ to $1.4 k_{res}^{-1}\sim 140\,d_i$.
These findings suggest, that both the damping and inverse cascade follow the linear analysis of two-wave interactions and their extension to wave spectra \cite{lee+73,volk+82}.

\subsubsection{Drift Speed in Stage IV}
\begin{figure}
    \centering
    \includegraphics[width=0.48\textwidth, trim=12 12 10 10, clip=true]{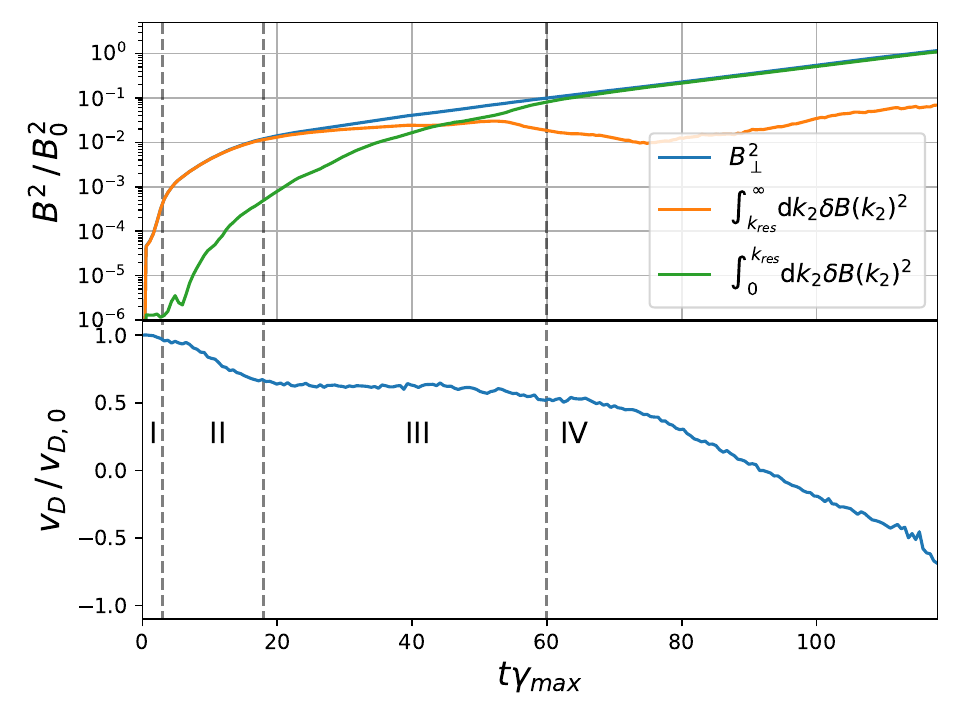}
    \caption{Top panel: Time evolution of the total perpendicular magnetic field (blue) and magnetic field on all scales smaller (orange) and larger (green) than the Larmor radius of CRs. Time is multiplied by the predicted growth rate of the instability.
    Bottom panel: Box-averaged CR drift velocity as a function of time.}
    \label{fig:caseII}
\end{figure}

\begin{figure}
    \centering
    \includegraphics[width=0.48\textwidth, trim=12 12 10 10, clip=true]{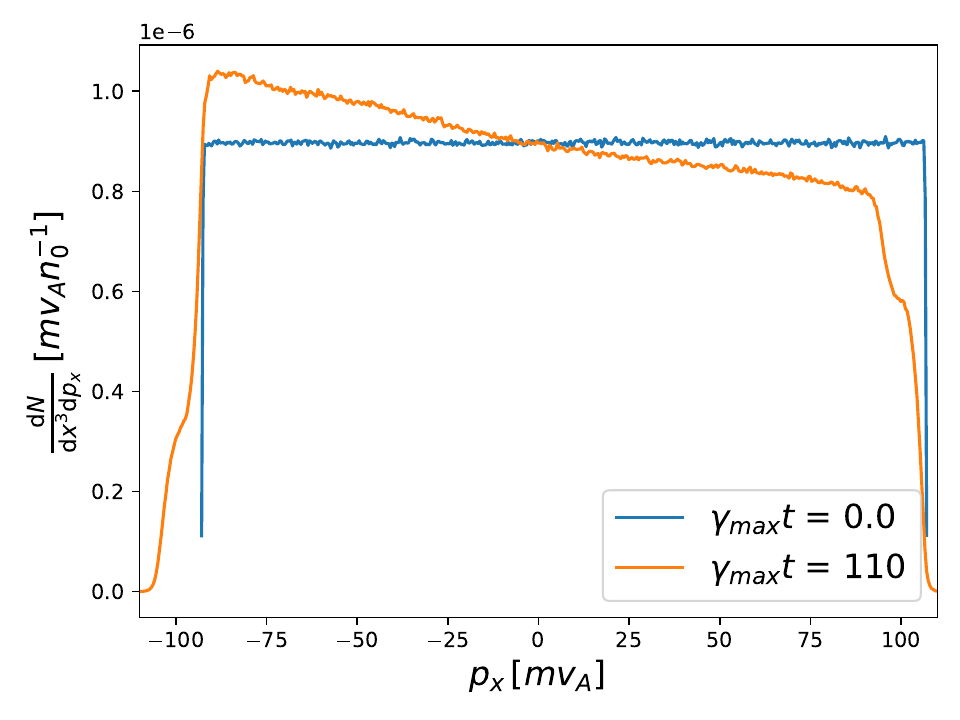}
    \caption{CR distribution function in $x$ momentum at the beginning and end of the simulation of case $\mathcal{B}$.}
    \label{fig:caseII_p1x1}
\end{figure}

The drift speed in Fig.~\ref{fig:caseII} shows also a peculiar behavior during Stage IV.
The scattering off the non-resonant waves leads to a reduction in drift speed that leads to an effectively negative drift speed.
The reason can be seen in Fig.~\ref{fig:caseII_p1x1}, showing the distribution of CRs as a function of $p_x$.
Initially, the distribution is flat with a small shift towards positive $p_x$ translating to the initial, positive drift speed.
At late times, the waves produced by the cascade scatter the CRs with positive $p_x$ more efficiently than the negative ones.
This effect leads to a pile-up in CRs with negative momenta while at the same time removing CRs out of the positive $p_x$ phasespace.
The result is an ever decreasing drift speed that eventually becomes negative.
The whole dynamics of the system are determined by highly nonlinear processes, as $\delta B/B_0$ approaches $1$, at which point we stopped the simulation, since no steady state is possible under these conditions, as discussed above.
The evolution of case $\mathcal{B}$ illustrated the importance of choosing the right boundary conditions corresponding to the physical picture one has in mind.
It further serves as a useful cross-check for the expectations based on the scalings of growth and damping rates.

\subsection{Case $\mathcal{C}$}
\begin{figure}
    \centering
    \includegraphics[width=0.48\textwidth, trim=5 10 10 10, clip=true]{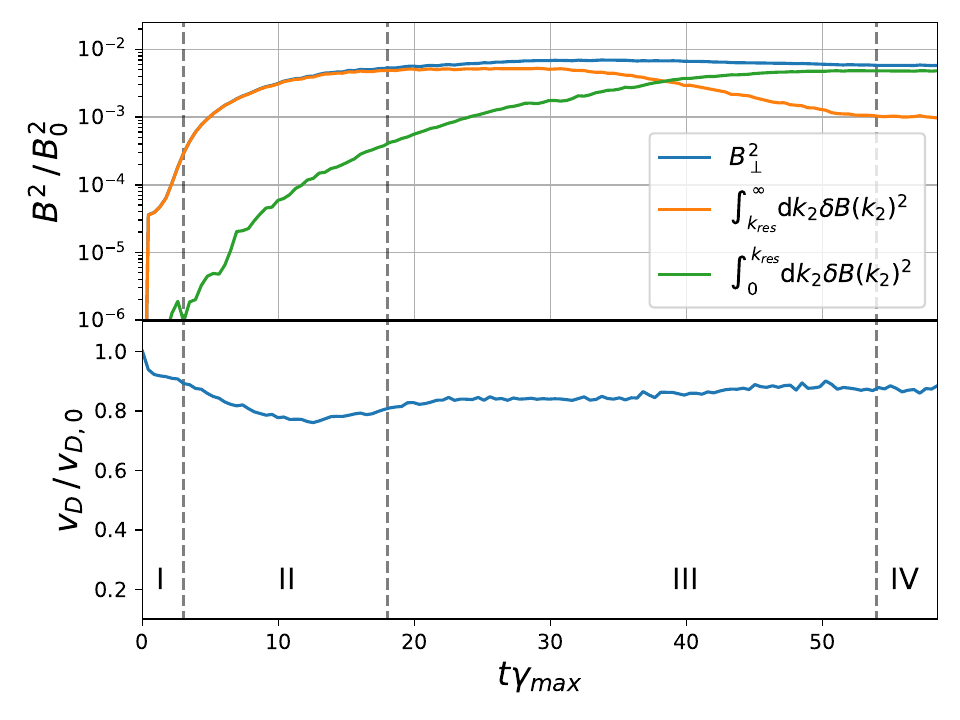}
    \caption{Top panel: Evolution of the transverse magnetic fields as in Fig.~\ref{fig:caseII}, but for case $\mathcal{C}$. 
    Note that in this case the growth rate and resonant wavenumber are a factor $3$ smaller than in the other cases due to the different CR population.
    Bottom panel: Box averaged CR drift velocity as a function of time normalized by its initial value $v_{D,0}=12\,v_A$. }
    \label{fig:caseIII}
\end{figure}

Case $\mathcal{C}$ is identical to case $\mathcal{A}$, except that the CRs are substituted with a population with higher energy and lower density inspired by the steeply falling spectrum of Galactic CRs. The new setup results in a larger growth time and resonant wavelength by a factor of $3$. To achieve a similar resolution in $k$-space around $k_{res}$, the box is a factor $3$ bigger than in case $\mathcal{A}$.
It is no surprise that the magnetic field and CR drift speed in Fig.~\ref{fig:caseIII}
follow the same evolutionary stages of pristine growth and subsequent scattering.
However, a fundamental difference is the strength of the saturated field.
The field only grows until $\delta B^2/B_0^2 \approx 0.005$, comparable with the magnetic turbulence expected in the Galactic halo.
The field saturates at lower values because the growth rate is a factor $3$ lower than in case $\mathcal{A}$. 
Hence, equality between damping and growth is expected already with magnetic fields on large scales that are a factor $\sim3$ less than before.
In fact, saturation of the resonant fields is achieved when the large-scale fields (green line in Fig.~\ref{fig:caseIII}) reach values of $\sim 4\times 10^{-4}$ instead of $\sim 10^{-3}$. 
Similarly as before, the large-scale modes are entirely generated by the inverse cascade on scales larger than the gyroradius of the CR particles.

Due to the larger simulation domain, the CR scattering length is resolved within our box even at such low magnetic field values.
The CRs scatter but their drift speed only changes by $25\,\%$, indicating that the field would grow to much larger values in the absence of damping.
Due to the larger box the Alfvén-passing time of the box is a factor $3$ larger than before. 
As expected, Stage IV is achieved roughly one Alfvén-passing time after the peak of the resonant fields. 
In this final stage, a steady state is achieved, visible by all fields and the drift speed becoming constant in time.
The overall behavior is in perfect agreement with case $\mathcal{A}$. We observe the saturation of the resonant fields at a damping rate that is compatible with the new CR population. 

\subsection{Case $\mathcal{D}$ and the Interplay of two CR Populations}
All our previous results are obtained for monoenergetic CR populations. In the Galaxy, CRs follow a power-law distribution over many orders of mangitude in energy. Hence, it is crucial to understand whether the presence of higher/lower energy CRs changes the above picture, since they might interact differently with the waves.
Therefore, in case $\mathcal{D}$, we take case $\mathcal{A}$ and add the CR population of the previous section.
The result allows us to study the interplay of the streaming instability and CRs with different momenta.
To facilitate the following discussion we divide the CRs into populations L (low energy) and H (high energy) and introduce the following names for the three regimes of the wave spectrum: waves resonant with population L ($k^{-1}\leq R_{L}$), waves resonant with population H ($k^{-1}\leq R_{H}$) and non-resonant modes ($k^{-1}\geq R_{H}$).

\begin{figure}
    \centering
    \includegraphics[width=0.48\textwidth, trim=12 12 10 10, clip=true]{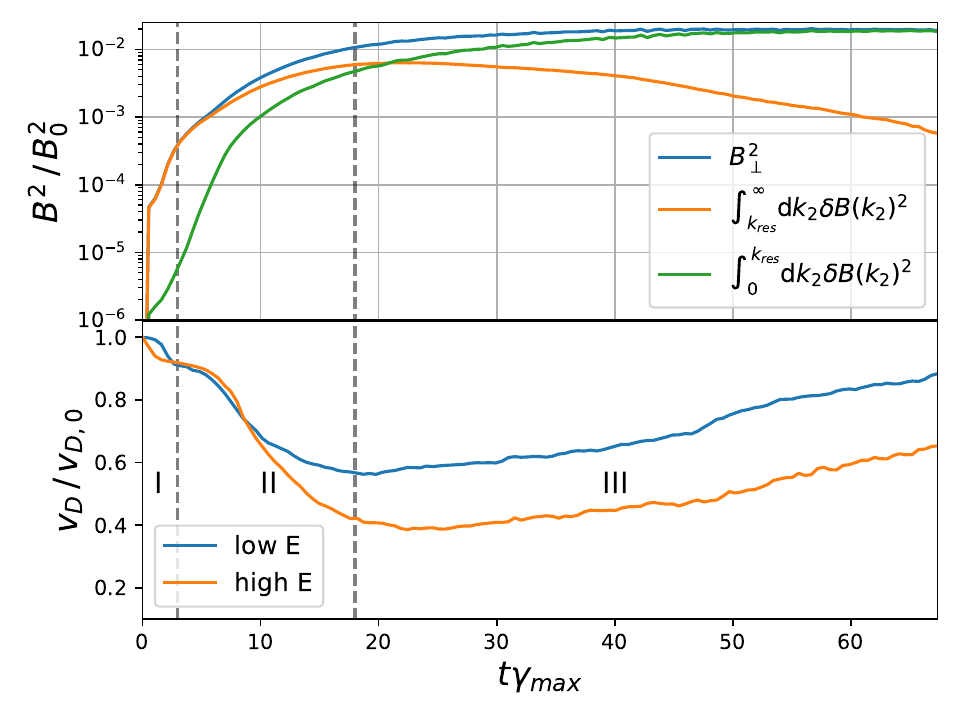}
    \caption{Same as Fig.~\ref{fig:caseII}, but for case $\mathcal{D}$. Since the drift speed is shown for two CR populations, the velocities are normalized to their individual initial drift speed, i.e., $4\,v_A$ and $12\,v_A$ respectively.}
    \label{fig:caseIV}
\end{figure}

High energy CRs with small pitch angles contribute to the growth of waves resonant with population L.
Therefore, the growth rate of these modes is larger than in case $\mathcal{A}$ and $\mathcal{B}$.
Following the standard argument of growth rate being equal to damping rate, we would expect the fields resonant with population L to saturate when a larger damping rate is achieved than in the previous cases.
This gives another separate, unique way to test the scaling of the damping rate.
If damping were to occur only due to interactions of waves close in wavenumber, we would expect the orange line in Fig.~\ref{fig:caseIV} to saturate at a slightly larger value than in the previous case. Intriguingly, this is not the case.
Instead, the total field resonant with population L is a factor of $\sim 2$ smaller than in cases $\mathcal{A}$ and $\mathcal{B}$.
This behavior underlines that the damping rate scales as in Eq.~\ref{eq:integral}.
The saturation in case $\mathcal{A}$ and $\mathcal{B}$ is achieved when the field on larger scales (green line) reaches values of $\delta B^2/B_0^2 \sim 0.002$.
In case $\mathcal{D}$, saturation occurs when the green line in Fig.~\ref{fig:caseIV} reaches $\delta B^2/B_0^2 \sim 0.003$ compatible with the slightly larger growth rate.
The reason why the fields resonant with population L saturate at a lower values is because the high-energy CRs generate magnetic fields on larger scales than population L ($R_{L} \leq k^{-1} \leq R_{H}$).
Hence, the critical value to compensate the growth of small-scale modes is achieved earlier in time, when they had less time to grow compared to cases $\mathcal{A}$ and $\mathcal{B}$.

Due to the stronger growth of larger scale waves, the equilibrium between waves leaving the box and newly generated waves is achieved at a larger value of the magnetic field in large-scale modes than in case $\mathcal{A}$. Hence, the waves resonant with population L are damped away entirely, as seen in Fig.~\ref{fig:caseIV}.
Note that also in case $\mathcal{A}$, we can observe the field on resonant scales decreasing over a short amount of time. 
However, in that case, the growth of large-scale modes is coupled with the power in small-scale modes due to the inverse cascade. Hence, reducing the magnetic field at small scales would indirectly also reduce the magnetic field at larger scales in case $\mathcal{A}$, an effect that is absent in case $\mathcal{D}$ due to the streaming instability excited by population H.

One can see that the green and blue lines in Fig.~\ref{fig:caseIV} flatten out, indicating that the growth of waves resonant with population H due to the streaming instability is saturating. However, the power in small-scale modes is still decreasing and, as a result, the drift speeds of both CR populations is still slightly increasing. 
At scales above $R_{H}$, we see again the inverse cascade developing, exactly as in case $\mathcal{A}$, just shifted to the largest wave number supported by streaming instabilities. Hence, we expect the simulation to evolve similarly to case $\mathcal{A}$.
Following the same reasoning as before, we expect the steady state to be achieved one Alfvén crossing time after the saturation of the streaming instability. Due to the larger required box size, the Alfvén crossing time would be a factor $\sim 3$ larger than in case $\mathcal{A}$.
Hence, the steady state would be achieved at $\gamma_{max}t \sim 140$, which is numerically very challenging to achieve, due to its high computational cost.

Interestingly, the power on scales resonant with population L is almost entirely damped away. Despite this difference compared to case $\mathcal{A}$, the drift speed of population L converges to a similar asymptotic value, indicating that non-resonant scattering takes over to account for the reduced effectiveness of resonant scattering.
Stronger non-resonant scattering can also explain why the transition from the minimum drift speed to its asymptotic value is much slower in case $\mathcal{D}$ despite the faster damping of resonant modes.
Instead of reaching the asymptotic value after $\sim 30 \gamma_{max}^{-1}$ as in case $\mathcal{A}$, it takes over twice as long in case $\mathcal{D}$.

The drift speed of population H is severely different than in case $\mathcal{C}$.
The high-energy particles are affected by the waves grown by population L by resonantly scattering with these waves below pitch angles $\mu\sim 0.3$.
Due to the wave power being larger than in case $\mathcal{C}$, the drift speed decreases considerably.
After the initial dip, it slowly increases again as the small-scale waves responsible for scattering are damped away, while the large-scale modes are growing to a level to maintain this level of scattering.
The drift speed of population H behaves almost identical to the one of population L. This indicates that both populations scatter off the same waves, further hinting towards efficient non-resonant scattering of population L.

The overall behavior, of more power on large scales, stronger damping, less resonant fields and a cascade above the largest scale $R_{H}$ is in good agreement with the predictions based on our previous simulation.
Furthermore, it confirms independently that the damping rate depends on the power on all scales larger than the resonant ones, rather only the resonant scales.

\section{Caveats and Limitations}
We now discuss some of the caveats of our numerical approach.
The first is the dimensionality of our simulations;
due to computational limitations, we can test this phenomenon only in quasi-1D boxes, as it requires extremely large boxes to achieve good resolution in $k$-space and long time scales due to the slow growth and cascade. 
We cannot test whether other effects occur when in higher dimensions, as is the case, e.g., for MHD turbulence \cite{muller+00}. 
The question of how inverse and direct cascades evolve in higher dimensionalities in kinetic simulations is still an open problem.

In the physical picture, the current of CRs passing through the Galactic halo is given solely by the rate of CR injection from their sources. The balance of damping and streaming instabilities then determines how this current is realized locally in terms of a CR density and drift speed in the halo. 
On the other hand, when injecting CRs in our simulations, we have to choose a density and drift speed as boundary conditions instead of just the rate of injection. The chosen values should mimic the CR distribution right outside of the box (in a nearby part of the halo).
In principle, once the CRs scatter off their self-generated turbulence, the density and drift speed could be very different than what is injected, which would contradict the physical situation we have in mind.
Hence, an important crosscheck is to ensure that the final drift speed and density are close to the injected ones.
In our simulations, the drift speed remains close to the initial one (within $10\,\%$). Hence, our simulation boxes should be representative of a location in the halo with CRs streaming through.

\section{Physical Implications}\label{sec:implications}
One of the main applications of CR streaming instabilities balanced by NLLD is CR transport in the Galactic halo.
Past studies have shown that the measured break in the B/C ratio \cite{dampe22,CALET22,ams21} can be interpreted as a transition of scattering regime, from scattering off self-generated (intrinsic) to scattering off extrinsic turbulence \cite{blasi+12,evoli+18b,aloisio+13}.
In these models the NLLD rate was calculated via Eq.~\ref{eq:resonant}, hence we now test how this picture would be modified by employing the correct formula (Eq.~\ref{eq:integral}).

As discussed in the previous section, the flux of CRs leaving the halo is balanced by the injection of CRs at the source \cite{aloisio+13}:
\begin{equation}
    D\partial_z f = \frac{\mathcal{R}_{\rm SN}}{\pi R_d^2} q(p),     
\end{equation}
with the rate of supernovae $\mathcal{R}_{\rm SN}$ and the radius of the Galactic disk $R_d$. The CR spectrum injected by a single supernova is normalized to the energy that ends up in CRs $E_{\rm SN,CR} = \int\mathrm{d}p^3 E(p) q(p)$.
Assuming a power law with $\gamma \neq4$, this results in 
\begin{equation}
    q(p) = \frac{E_{\rm SN,CR} (\gamma-4)}{4 \pi c p_0^4} \left(\frac{p}{p_0}\right)^{-\gamma}
\end{equation}
with $p_0\approx 1\,$GeV being the smallest momentum.
The growth rate of the resonant streaming instability can be written as \cite{skilling75c}:
\begin{align}
    \Gamma_{\rm CR} =& \frac{16 \pi^2}{3} \frac{v_A}{k\delta B(k)^2} p^4 v \partial_z f = 4 \pi \Omega_0 \frac{p^3 D\partial_z f}{n_0 v_A}\\
    =& 4\pi \frac{\Omega_0}{n_0 v_A} \frac{\mathcal{R}_{\rm SN}}{\pi R_d^2} p^3 q(p)
\end{align}
where we have used the self-generated diffusion coefficient \citep{skilling75a}:
\begin{equation}\label{eq:diffusion}
    D = \frac{v}{3} \frac{pc}{eB_0}\frac{B_0^2}{k\delta B(k)^2}\,.
\end{equation}
The damping rate should be given by the integral over all power at larger scales. The damping of smaller-scale modes arising only from the extrinsic turbulence injected at large scales is \cite{lagage+83b,chernyshov+22}:
\begin{equation}
\Gamma_{\rm NLLD} = g kv_A \int_0^k\mathrm{d}k' \frac{\delta B(k')^2}{B_0^2} = g \frac{v_A}{r_0} \eta_B \left(\frac{p}{p_0}\right)^{-1}
\end{equation}
with $\eta_B$ defined as the integral, following Ref.~\cite{aloisio+13}, and the prefactor $g$ being a function of $\beta$ and of order of a few. Note that $\eta_B$ is independent of $k$ because we assume the extrinsic turbulence to be injected at scales much larger than any resonant scale of CRs.
Since we are interested in damping at resonant scales, $k^{-1}$ was replaced by the gyroradius $r_L = r_0 \frac{p}{p_0}$. 
It is important to stress that strictly speaking these considerations  apply only to the case of extrinsic \textit{Alfv\'enic} turbulence. 
We are unable to comment, at this point, on what would be the contribution of non-Alfv\'enic modes on large scales to the NLLD of Alfv\'en waves generated by low energy CRs.

In order to self-confine in the presence of this external turbulence, the growth of the instability must overcome this damping.
Putting everything together, we derive the following constraint for the momenta at which self-confinement is efficient:
\begin{align}\label{eq:selfconf}
    \frac{p}{p_0} \leq& \left( \frac{4 \pi}{B^2 c} \frac{\mathcal{R}_{\rm SN} E_{\rm SN, CR}}{\pi R_d^2} \frac{\gamma-4}{g \eta_B} \right)^{\frac{1}{\gamma-4}} \nonumber \\
    \approx& \left( 0.002 \frac{\gamma-4}{g\eta_B} \right)^{\frac{1}{\gamma-4}},
\end{align}
where we assumed $B_0\approx 3\,\mu$G, $\mathcal{R}_{\rm SN}= \frac{1}{30}\,$yr$^{-1}$, $R_d\approx 10\,$kpc and $E_{\rm SN,CR}\approx 10^{50}\,$erg.
Gamma-ray observations of supernova remnants and models of diffusive shock acceleration suggest spectra with $\gamma\sim 4.2-4.4$ \cite{caprioli11, caprioli12, caprioli+20}.
Even using $g=1$ and $\gamma=4.4$, CRs can self-confine above $1\,$GeV only if $\eta_B\lesssim 8\times 10^{-4}$.
Note that in the case of $\gamma=4$, the condition becomes independent of $p$, meaning that self-confinement is either efficient or inefficient for all values of the momentum $p$. By using the right normalization of $q(p)$ for this case, we obtain $\eta_B\lesssim 10^{-4}$ for self-confinement to work.

In the physical picture where the observed hardening in B/C is interpreted as a transition of scattering off self-generated to extrinsic turbulence \cite{blasi+12,aloisio+13,evoli+18b}, $\eta_B$ can be estimated by assuming a power-law spectrum for the extrinsic turbulence 
\begin{equation}
    \frac{k\delta B(k)^2}{B_0^2} = \eta_B (s-1) \left(\frac{k}{k_0}\right)^{1-s}
\end{equation}
with an injection scale $L_0=1/k_0$
and comparing the resulting diffusion coefficient to inferred values at $\sim \,$TeV energies.
Using $D(1\,\mathrm{TeV})\approx3\times10^{29}\,$cm$^2$/s, this estimate gives $\eta_B\approx 0.02 \left(\frac{L_0}{10\,\mathrm{pc}}\right)^{2/3} \left(\frac{p}{1\,\mathrm{TeV}}\right)^{1/3}$
for a Kolmogorov spectrum, comparable with $\eta_B\approx0.05$ used in previous studies \cite{aloisio+13}.
These values are much larger than the threshold for self-confinement to work at GeV energies derived in Equation \ref{eq:selfconf}.
Hence, the interpretation of the hardening as due to a transition between different scattering regimes faces serious problems with the correct prescription of NLLD.
The presence of even very small perturbations on large scales potentially challenges CR self-confinement at GeV energies in the Galaxy. 

The considerations above suffer from similar caveats as the original idea of the transition of scattering regimes \cite{blasi+12}: 
i) The extrinsic turbulence might not be Alfvénic, in which case, it would not partake in NLLD and resonant modes might still be excited. 
ii) If it is Alfvénic, then the cascade could be anisotropic in $k$-space and have little power in $k_{\parallel}$ \cite{goldreich+95}, the direction that matters for NLLD. 
In this case scattering might be due to intermittency \cite{lemoine+23,kempski+25}, though in this case the resulting diffusion coefficient is not well understood.
Preliminary estimates suggest that intermittency may be negligible at $E\lesssim 10\,$TeV where we have measurements of B/C, and diffusion at low energies would still need to rely on other processes, such as self-generation and NLLD.

Whether there is no extrinsic large-scale turbulence in our Galaxy and all fields are self-generated, as suggested in Ref.~\cite{chernyshov+24,chernyshov+22}, or whether the previous picture \cite{blasi+12,evoli+18b} can be recovered with some modifications will be investigated in a future work together with the effect of the inverse cascade, neglected here to point out the zero-th order effects.
Nonetheless, this simple estimate highlights the importance of using the correct damping formula.

\section{Conclusions}
\label{sec:conclusions}
In many different environments, such as the Galaxy \cite{chernyshov+24,evoli+18b} or near sources \cite{nava+16,nava+19}, CRs self-confine, which means that they diffuse in self-generated magnetic fields.
In the hot ionized medium, the growth of these fields is thought to be balanced by NLLD.
Here, we present a detailed analysis of this damping with hybrid PIC simulations and present inferred scalings of the damping rate as well as limitations of the numerical approach.

In our simulations, CRs are injected into a periodic box of background gas and generate the resonant streaming instability.
A steady state is achieved in a box that the Alfvén waves can leave, as presented in Ref.~\cite{schroer+25}.
Contrary to the common approximation \cite{wiener+13,recchia+16,recchia+22,nava+19}, we find that the damping rate at a given scale depends in fact on the power of magnetic fields on larger scales.
An inverse cascade develops, which produces magnetic fields on scales non-resonant with any particles in our simulation. Such a shift from magnetic fields on small to large scales can be explained with the linear theory of NLLD where a wave gains energy when interacting with another wave of smaller wavelength \cite{lee+73}.

To further test the occurrence of this phenomenon, we present here several numerical tests with the goal of outlining the importance of this phenomenon for Galactic CR transport.
First, we test whether the cascade does naturally stop after it is initiated. This is an important check, since the analytical prediction suggests that an equilibrium is impossible as long as small-scale modes are driven by, e.g., CRs.
For this test, we perform the same simulation but make the box periodic for the waves. In this way, magnetic energy cannot escape the system and a balance can only be achieved by a balance of growth and damping.
At first, a metastable phase is reached, in which CRs diffuse with a constant drift speed. However, the inverse cascade moves most of the  power to large scales and the total magnetic field keeps growing, proving that a steady state cannot be achieved without another way to dissipate the fields at large scales.
Our simulations further show a peaked spectrum of waves. We show that such a spectrum is compatible with the linear analysis of two-wave interactions \cite{lee+73} and the peak can be interpreted as the balance of wave growth and damping at a given $k_0$ that slowly decreases with time.

To apply our results to phenomenological studies, we extend our simulations beyond monoenergetic CR populations by adding a second populations of higher-energy CRs. 
We find that these CRs generate waves via the resonant streaming instability and the inverse cascade populates scales even larger than their resonant scale. The smaller-scale waves get damped stronger than without these high-energy CRs, confirming that damping depends on all the power contained on larger scales.
All of these findings are consistent with the expectations based on the simulation with a single CR species and suggest that they can be applied to CRs in the Galaxy.

An inverse cascade generating fields on large scales could have profound implications for CR transport in the Galaxy.
A phenomenological description of this phenomenon is beyond the scope of this article and will be discussed in a future work.
Here we limit ourselves in estimating the effect of the presence of large-scale Alfv\'enic turbulence on CR self-confinement in the Galaxy. Using the correct formula for NLLD, we find that CRs are unable to self-confine even at GeV energies if Alfvènic modes  are present on large scales with $\delta B^2/B_0^2\gtrsim 10^{-4}$.
This shows the importance of using the proper damping rate and implies that modifications to the previous picture are needed in order to retain the interpretation of the hardening.

\acknowledgements
Simulations were performed on computational resources provided by the University of Chicago Research Computing Center and on  TACC's Stampede3 through ACCESS Maximize allocation PHY240042.
We thank Elena Amato, Anatoly Spitkovsky, Philipp Kempski, Matthew Kunz, Christopher Pfrommer, Mohamad Shalaby, Ellen Zweibel, and Rouven Lemmerz for interesting and stimulating discussions, some of which took place at Aspen Center for Physics, which is supported by National Science Foundation grant PHY-2210452. 
This work of D.C. was partially supported by NASA grant 80NSSC18K1726, NSF grants AST-2510951 and AST-2308021, and by NSF grant PHY-2309135 to the Kavli Institute for Theoretical Physics.
The work of P.B. was partially supported by the European Union – NextGenerationEU RRF M4C2 1.1 under grant PRIN-MUR 2022TJW4EJ and by the European Union - NextGenerationEU under the MUR National Innovation Ecosystem grant  ECS00000041 - VITALITY/ASTRA - CUP D13C21000430001.

\bibliographystyle{apsrev4-2}
%

\end{document}